\documentclass[a4paper,aps,nofootinbib,twocolumn]{revtex4}
\usepackage[english]{babel}
\usepackage{amssymb,amsthm}

\begin{document}

\renewcommand{\c}[1]{{\cal{#1}}}
\def\H{{\c H}}
\def\K{{\c K}}
\def\R{{\c R}}
\def\BH{{\c {BH}}}

\title{Black holes in full quantum gravity}
   \author{Kirill Krasnov$\,{}^a$, Carlo Rovelli$\,{}^b$}
\affiliation{${}^a\,$Mathematical Sciences, University of Nottingham, Nottingham, NG7 2RD, UK\\
%}
  % \author{Carlo Rovelli}      
     %\affiliation{
     ${}^b\,$Centre de Physique Th\'eorique de Luminy\footnote{Unit\'e mixte de recherche (UMR 6207) du CNRS et des Universit\'es de Provence (Aix-Marseille I), de la M\'editerran\'ee (Aix-Marseille II) et du Sud (Toulon-Var); laboratoire affili\'e \`a la FRUMAM (FR 2291).}, Case 907, F-13288 Marseille, EU}   
\date{\today}
  
\begin{abstract}
Quantum black holes have been studied extensively in quantum gravity and 
string theory, using various semiclassical or background dependent approaches.  
We explore the possibility of studying black holes in the full non-perturbative 
quantum theory, without recurring to semiclassical considerations, and in the context of loop 
quantum gravity.  We propose a definition of a quantum black hole as the collection of the quantum 
degrees of freedom that do not influence observables at infinity.  From this definition, 
it follows that for an observer at infinity a black hole is described by an ${\rm SU}(2)$ 
intertwining operator.  The dimension of the Hilbert space of such intertwiners grows
exponentially with the horizon area. These considerations shed some light on the
physical nature of the microstates contributing to the black hole entropy. In particular, 
it can be seen that the microstates being counted for the entropy have the
interpretation of describing different horizon shapes. The space of black
hole microstates described here is related to the one arrived at
recently by Engle, Noui and Perez, and sometime ago by Smolin, but 
obtained here directly within the full quantum theory.

\end{abstract}
\maketitle

\section{Introduction}

Considerable progress  has been obtained in understanding the microphysics 
of black hole entropy using loop quantum gravity \cite{bhe}, following the
first pioneering works started in the late nineties \cite{BHold,lee}.  So far, 
however, the description of black holes  has relied on some mixture of quantum theory 
and classical analysis of black hole geometry: for instance, one can characterize a 
black hole classically \cite{IH}, and then quantize the part of the classical-theory 
phase space that contains the black hole.  Is it possible, instead, to describe black holes 
entirely within the non perturbative quantum theory of spacetime \cite{book,Rovelli:1989za,lqg}? 

In this paper we suggest a direction for answering this question. We 
propose a simple definition of a quantum black hole within the 
full quantum loop theory, as a region of a spin network which is not ``visible"
from infinity. This is in the same spirit of the global analysis that is possible 
in classical general relativity, where properties of horizons and black holes can be
obtained by studying their implicit definition, even without being
able to solve the equations of motion and writing the metric explicitly \cite{HE}.

We use this definition to study how black holes 
are characterized quantum mechanically and find that they can be described 
by ${\rm SU}(2)$ intertwining operators.  The Hilbert space of such operators
is intimately related with the space of states of ${\rm SU}(2)$ Chern-Simon (CS) theory on a punctured surface: the two spaces have the same dimension
if the CS level is high enough, and the former arises from the later in the limit when 
the level is taken to infinity.  The space of ${\rm SU}(2)$ Chern-Simon states as
describing the black hole quantum microstates was obtained recently by 
Engle, Noui and Perez \cite{ENP} developing the work of Ashtekar, Baez, Corichi and 
Krasnov \cite{bhe}, that is, using
a semiclassical approach where a boundary condition is imposed on the classical
theory before quantization. It is also the space obtained a time ago 
by Smolin \cite{lee}, using related semiclassical considerations.
This paper shows that a Hilbert space closely related to that of ${\rm SU}(2)$ CS theory
emerges from a very natural  characterization of quantum black holes in the full theory.

Furthermore, we argue that, contrary to what is often assumed, these states
are distinguishable, in an appropriate sense, by measurements outside the
hole.  They are related to fluctuations of the extrinsic geometry of the 
black horizon. 

\section{Preliminaries}

We refer to \cite{book} for an introduction to loop quantum gravity 
and for notation.  we briefly recall here only the elements of the 
theory that are needed below.  The kinematical 
Hilbert space $\c K_{\rm aux}$ on which the theory is defined 
admits a linear subspace $\K_{\Gamma}$ for each graph
$\Gamma$ embedded in a three dimensional manifold 
$\Sigma$.\footnote{Technically, $\K_{\rm aux}$ is a projective limit
built from the spaces $\K_{\Gamma}$ \cite{book}.}
The space $\K_{\Gamma}$ is the Hilbert space of an ${\rm SU}(2)$
lattice gauge theory on $\Gamma$. That is, $\K_{\Gamma}=
L_2[{\rm SU}(2)^{L}]$, where $L$ is the number
of links in $\Gamma$. The theory is invariant under local
${\rm SU}(2)$ transformations.  The ${\rm SU}(2)$ gauge invariant states
live in the subspace $\H_{\Gamma}=L_2[{\rm SU}(2)^{L}/SU(2)^{N}]\subset \K_{\Gamma}$,
where $N$ is the number of nodes in $\Gamma$ and the action
of the gauge transformations is $\psi(U_l)\to\psi(V_{i(l)}^{-1}U_lV_{f(l)})$
where $i(l)$ and $f(l)$ are the initial and final nodes of the link $l$ and
$U,V\in {\rm SU}(2)$.  Peter-Weyl theorem implies that $\K_{\Gamma}$ decomposes as 
$$
\K_{\Gamma}=\oplus_{j_l}\otimes_n(\otimes_{a}H_{j_a}),
$$ 
where $l$ are the links in $\Gamma$, $n$ the nodes, $a$ are the links adjacent to the node $n$ and
$H_{j}$ is the Hilbert space of the spin $j$ irreducible representation of ${\rm SU}(2)$.
Then 
$$
\H_{\Gamma}=\oplus_{j_l}\otimes_n({\rm Inv}[\otimes_{a}H_{j_a}]),
$$
where the space $\H_n={\rm Inv}[\otimes_{a}H_{j_a}]$ is that of ${\rm SU}(2)$-invariant
tensors -- intertwiners -- of the node $n$. 
The ${\rm SU}(2)$ gauge invariant state space of the theory  $\H_{\rm aux}$ is composed by 
(is the projective limit of) all the $\H_\Gamma$ spaces.  Choosing a basis $i_n$ in each 
$\H_n$ of each  $\H_\Gamma$, we obtain the spin network basis 
$|S\rangle\equiv|\Gamma,j_{l},i_{n}\rangle$ for the ${\rm SU}(2)$-gauge 
invariant states of the theory.  

$\H_{\rm aux}$ carries a unitary representation of the group 
Diff${}_{\Sigma}$ of the diffeomorphisms of $\Sigma$. This allows us to 
define and solve the 3d diffeomorphism gauge of the theory.  The Hilbert 
space of diff-invariant states $\H_{\rm diff}$, admits 
an orthonormal basis $|s\rangle$, where $s$ (referred to as s-knot or simply a spin network)
is an equivalence class of embedded spin networks $S$ under diffeomorphisms. There exists a projection operator 
$\pi: \H_{\rm aux}\to \H_{\rm diff}$, which can be intuitively viewed as the 
exponentiation of the quantum diffeomorphism constraint 
operator.  An operator $O$ on $\H_{\rm aux}$ is diffeomorphism invariant if 
there exist an operator $O_{\rm diff}$ on $\H_{\rm diff}$ such that $\pi O= 
O_{\rm diff}\pi$.

The first assumption on which our result relies is that there exists a projection 
operator $P: \H_{\rm diff}\to \H_{\rm phys}$ implementing 
the dynamics. Here $\H_{\rm phys}$ is the space of 
the physical states (``the solutions of the Wheeler DeWitt 
equation''). An operator $O$ on $\H_{\rm diff}$ is gauge invariant (i.e., 
is a ``physical operator'') if there exist an operator $O_{\rm phys}$ on 
$\H_{\rm phys}$ such that $P O_{\rm diff}= O_{\rm phys}P$.  The set of operators 
$O$ on $\H_{\rm aux}$ such that $P \pi O = O_{\rm phys}P\pi$ form the gauge 
invariant observable algebra $\cal A$.  There are various attempts to 
define the quantum Hamiltonian constraint \cite{book} or directly 
the operator $P$ \cite{SpFoams}, but the argument we present here 
depends only on the existence of $P$, and
not on its specific form.

Our second assumption concerns asymptotic flatness.  Most of 
the work in loop quantum gravity has so far assumed 3d
physical space to be compact.  Here, we assume that a suitable 
extension of the theory to the asymptotically flat case 
exists.  In particular, in what follows we make use of the notion of 
an asymptotic observer.  

Consider classical general relativity in the asymptotically flat case.  
Let ${\cal C}$ be the space of the initial (Cauchy) data of the theory on a 
spacelike surface $\Sigma$.  Using the evolution equations, we can 
compute the value of the gravitational field at any spacetime point in the 
future of $\Sigma$.  In particular, given the initial data we can compute the value of the 
gravitational field at future null infinity. Therefore observables at null 
infinity are functions on the initial-data phase space. They are non-local
and very complicated functions, since writing them explicitly requires 
solving the equations of motion, but they are nevertheless implicitly
well-defined.  Let $O$ be one such observable 
quantity at future null infinity.   

In the quantum theory, the space ${\cal C}$ of the initial data is promoted to a state 
space $\H_{\rm aux}$, and functions on ${\cal C}$ are promoted to operators 
on $\H_{\rm aux}$.  Then to every observable $O$ at 
future null infinity there is a corresponding (Heisenberg) operator $\hat{O}$ on $\H_{\rm aux}$.  
These operators define the algebra ${\cal A}_{\infty}$, which is a 
subalgebra of the algebra of physical operators.  
Observables at null infinity are gauge invariant, because they do not 
depend on arbitrary lapses or shifts of $\Sigma$.  Therefore the 
operators $\hat{O}$ must commute with the projection operators $\pi$ and 
$P$, and be well-defined on $\H_{\rm phys}$.  Our assumptions above are thus  
equivalent to an assumption that the algebra $\cal A$ of gauge invariant 
operators contains a subalgebra ${\cal A}_{\infty}$ of operators $\hat O$, that 
corresponds to all possible observations of the gravitational 
field that can be made at future null infinity. What follows depends 
on the existence, not on the explicit form of this algebra. 

\section{Definition of a quantum black hole}

Since we are working in the canonical formalism, we need 
a notion of a black hole at a spacelike surface. 
In classical general relativity, a black hole is a region
of space which is outside the past of future null infinity. 
This can be formulated as follows. Consider a spacelike
surface $\Sigma$, and initial data $c$ on $\Sigma$.
Then a region $\c R$ of $\Sigma$ is inside a black hole
if all observables at null-infinity have the same value on
$c$ and on any other initial data $c'$ that are the same 
as $c$ outside $\R$.  Of course, to determine explicitly
if a certain region is or isn't inside a black hole, is a nontrivial
task (as numerical relativity people know well), since one must 
in principle evolve the data to infinity in order to find out;
nevertheless (again, as numerical relativity people know well), 
the notion is well defined.  

It is useful to define the {\em external} and {\em
internal} geometry of a (open) spacial region ${\cal R}$ as follows:
The {\em external} geometry of the region is ensemble
of the properties of the geometry that can be measured by 
local observables which are not in the region. 
Examples include the intrinsic (e.g. area) as well as extrinsic (e.g. extrinsic curvature) 
geometry of the boundary of the region. The {\em internal} geometry of the region is the ensemble
of properties of the region that can be measured only 
by local observables in the interior of the region. An example
is the volume of the region. Thus, a region ${\cal R}$ is inside
a black hole if all observables at null infinity are independent
of the internal geometry of $\R$; namely if they have the same
value on any other initial data that have the same external
geometry.  Let us try to capture the same idea in the quantum theory. 

Consider a spin network state $|S\rangle$. It defines a state $S:{\cal 
A}\to \mathbb{C}$  over the algebra $\cal A$ of the observables by 
$S(O)=\langle S|O |S \rangle$.  
Consider a region ${\cal R}$ in $\Sigma$. 
Denote ${\cal A}_{\cal R}$ the subalgebra of the observable
algebra $\cal A$ formed by all \emph{local} observables with 
support in $\cal R$.   Call $S_{{\cal R}}$ the restriction of $S$
to  ${\cal A}_{\cal R}$.  Similarly, let  $S_{\infty}$ be the restriction of $S$ to 
$A_{\infty}$. 
Given a spin network $S$, let us say that an open region $\cal R$ is a 
``hidden region" in the quantum state $|S\rangle$ iff 
\begin{equation}
S_{\overline{\R}}=S'_{\overline{\R}} \hspace{2em} \Rightarrow
\hspace{2em}  S_\infty = S'_\infty, 
\end{equation} 
where $\overline{\R}$ is the complement of $\R$. The algebra $A_{\overline{\R}}$ is formed 
by all \emph{local} observables that have \emph{no} support on $\R$. These 
are the observables that read the ``outside geometry" of $\R$.  
Therefore this definition captures precisely the notion of a region that
does not affect infinity: any other state $|S'\rangle$ which is equal to
$|S\rangle$ \emph{outside} the hidden region (meaning: that is indistinguishable from 
$|S\rangle$ by means of local measurements outside the hidden region) is also indistinguishable from
$|S\rangle$ when observed at infinity. That is: the part of $|S\rangle$
inside the hidden region does not affect the future infinity. 

Let us now call the maximal hidden region of a state $|S\rangle$ a ``black hole" 
region and denote it as ${\cal BH}(S)$. A spin network with a black hole splits into two parts: 
we call a ``quantum black hole" the portion of $S$ inside the hole, that is, the open graph 
$\Gamma_{\cal BH} := \Gamma\cap{\cal BH}(S)$ with its colourings, and denote it
$S_{\cal BH}:=(\Gamma_{\cal BH},j_{l_{\cal BH}},i_{n_{\cal BH}})$. For simplicity, we 
consider here only the situation when the quantum black hole $S_{\cal BH}$ is connected.  
We call $S_{\rm ext}:=(\Gamma_{\rm ext},j_{l_{\rm ex}},i_{n_{\rm ex}})$, the rest of the spin network,
that is, the open graph $\Gamma\cap\overline{{\cal BH}(S)}$, with 
its colourings.  Let us call ``internal black hole geometry" the set of quantum numbers
$S_{\cal BH}:=(\Gamma_{\cal BH},j_{l_{\cal BH}},i_{n_{\cal BH}})$ and
``external geometry" the set of quantum numbers
$S_{\rm ext}:=(\Gamma_{\cal BH},j_{l_{\rm ex}},i_{n_{\rm ex}})$. 

Since knowledge of the data at a node includes the knowledge about the links that arrive to this node,
all links that are bounded by nodes in $S_{\rm ext}$ are also in $S_{\rm ext}$. Then the 
links of $S_{\rm ext}$ split into two groups: those that are bounded by two external nodes and those that 
are bounded by a node in  $S_{\rm ext}$ and a node in $S_{\cal BH}$.  It is natural to
call these second kind ``horizon links". They form the open legs of the graph of $S_{\rm ext}$.  
Pictorially, they are the links that puncture the horizon of the black hole.  

It is important to observe that all definitions above are given in terms of a spin network $S$. 
The black hole region is only defined as a part of the graph of $S$, and in particular the horizon 
is only defined as a collection of links that separate $S$ into an external and an internal component.  
In other words, a quantum black hole defined in this way is not a sharp surface in the manifold $\Sigma$: 
it is only a split of a spin network.   A consequence is that the notion is immediately 
3-dimensionally diffeomorphism-invariant, and thus comes down to $\H_{\rm diff}$. 

In the next section we study properties of a quantum black hole just defined. 

\section{Observability and entropy}

The split $S\to (S_{\cal BH}, S_{\rm ext})$ between the internal and
external part of the black hole determines a split in the Hilbert space $\H_\Gamma$,
where $\Gamma$ is the graph of $S$. Indeed we can write  $\H=\H_{\cal BH}\oplus \H_{\rm ext}$, where
$$
\H_{\rm ext}=\oplus_{j_{l_{\rm ext}}}\otimes_{n_{\rm ext}}(\otimes_{a}H_{j_a}),
$$ 
and
$$
\H_{\cal BH}=\oplus_{j_{l_{\cal BH}}}\otimes_{n_{\cal BH}}(\otimes_{a}H_{j_a}),
$$ 
Since states in both spaces live on open graphs, they transform non-trivially under
local ${\rm SU}(2)$ gauge transformation.  If we label with an integer $p=1,...,P$ the horizon 
links, the states in $\H_{\cal BH}$ and $\H_{\rm ext}$ live in a representation of ${\rm SU}(2)^P$ 
with spin $\{j_p\}$, namely in $\H_{\rm horizon}=\otimes_p \H_{j_p}$. In other words, the states of 
both spaces have free magnetic indices where the graph $\Gamma$ has been cut.

Consider two states $|S\rangle$ and $|S'\rangle$. Let us say that they are ``equivalent" if 
$S_{\rm ext} = S'_{\rm ext}$, that is, if they are indistinguishable by measurements outside the 
black hole region.   Denote the corresponding equivalence classes by $|[S]\rangle$.  Because of 
the very nature of the horizon, for all observers that do not enter the horizon, a state containing 
a black hole is effectively described by the class $|[S]\rangle$.  

A crucial observation is now the following.  One may be tempted to deduce from the above considerations 
that the states  $|[S]\rangle$ are fully determined by the external geometry of the hole; namely by 
the quantum numbers $S_{\rm ext}=(\Gamma_{\cal BH},j_{l_{\rm ex}},i_{n_{\rm ex}})$.  But this is not the case. 
A state $|[S]\rangle$ is determined by \emph{more} degrees of freedom than those characterizing 
its outside geometry.  

To see this, consider an operator defined as follows.  Let $T_\alpha^{ab}(x,y)$ be the ``two-hand" 
grasping operator in terms of which loop quantum gravity was initially defined \cite{Rovelli:1989za}. 
This is the operator $T^{ab}[\alpha](x,y)={\rm tr}[E^a(x) U_{\alpha_1} E^b(y) U_{\alpha_2}]$, 
where $E^a(x)$ is the Ashtekar electric field, $U$ is the holonomy of the Ashtekar connection 
and $\alpha_1$ and $\alpha_1$ are two lines connecting $x$ and $y$. 
Let $T_{pp'}=\int_{\Sigma_p} d^2 x\int_{\Sigma_{p'}} d^2 y \ n_a(x)n_b(y)T^{ab}[\alpha](x,y)$, 
where  $\Sigma_p$ is a small surface punctured by the link $p$ and $n_a$ its normal. A moment of 
reflection shows that this operator has support \emph{outside} the black hole.  However, it 
reads properties of $|S\rangle$ that depend on features of the spin network $S$ \emph{inside} 
the black hole.   This is easily seen by acting with this operator on two links $p$ and $p'$ bounded 
by a node $n$ that is inside a BH; the action of the operator depends on the intertwiner at $n$. 

This shows that observables in the outside region \emph {can} read some features of $|S\rangle$ which 
are not captured by the quantum numbers $S_{\rm ext}=(\Gamma_{l_{\rm ex}},j_{l_{\rm ex}},i_{n_{\rm ex}})$. 
In other words, the ``external geometry", defined as what can be observed by observers with support 
outside the hole, is more rich that the ``outside geometry", defined by 
$(\Gamma_{l_{\rm ex}},j_{l_{\rm ex}},i_{n_{\rm ex}})$.  Indeed, to help intuition, notice 
that even a change of a intertwiner ``deep inside" $\Gamma_{\cal BH}$ can be detected by the 
observable $T_{pp'}$.   What are thus these additional degrees of freedom?

A moment of reflection shows that the additional degrees of freedom that can be observed  
by external observers are completely captured as follows. We have seen that a state in 
$\H_{\cal BH}$ transforms as a vector in $\H_{\rm horizon}=\otimes_p \H_{j_p}$. The operators $E^a(x)$ 
act as ${\rm SU}(2)$ generators on each $\H_{j_p}$.  The ${\rm SU}(2)$-invariance implies that
only the globally ${\rm SU}(2)$ gauge invariant subspace of this space is physically relevant. 
Therefore the degrees of freedom that can be read out by observables outside the hole, and are 
not captured by  $S_{\rm ext}:=(\Gamma_{\cal BH},j_{l_{\rm ex}},i_{n_{\rm ex}})$ are entirely 
determined by the state space 
$$
\H_{\rm horizon}={\rm Inv}[\otimes_p \H_{j_p}],
$$  
where the operator $T_{pp'}$ acts as $T_{pp'}\sim J_p^iJ_{p'}^i$, where $J^i_p, i=1,2,3$ are 
the ${\rm SU}(2)$ generators in $H_{j_p}$.  Thus, we conclude that 
$$ 
|[S]\rangle = |I_{\BH}, \Gamma_{\rm ext},j_{l_{\rm ex}},i_{n_{\rm ex}}\rangle
$$
where $I_{\BH}\in \H_{\rm horizon}$ is a single intertwiner.   
In other words, from the point of view of the outside observer, a black hole behaves as a 
(possibly gigantic) single intertwiner, which intertwines all the links puncturing its horizon. 
 
Suppose now that we are in a statistical mechanical context and want to count the number of 
states subject to given conditions. Suppose that we know the outside geometry (area) of the black hole 
horizon.  Then we must associate to the black hole an entropy equal to the (logarithm of the) 
number of states compatible with this outside geometry.  This number is given by the dimension 
of the Hilbert space $\H_{\rm horizon}$:
\begin{equation}
N={\rm dim}\ \H_{\rm horizon}={\rm dim}\ {\rm Inv}[\otimes_p \H_{j_p}].
\label{N}
\end{equation}
This dimension is given by the classical formula
\begin{equation}
N=\frac{2}{\pi}\int_0^\pi d\theta \,
\sin^2(\theta/2) \, 
\prod_p \chi^{j_p}(\theta)
\end{equation}
where $\chi^j(\theta)= \sin{((j+1/2)\theta)}/\sin{(\theta/2)}$ are the ${\rm SU}(2)$ characters.
For a large number of punctures this goes as $N\sim \prod_p (2j_p+1)$, which grows exponentially as
a function of the BH horizon area.

The space $\H_{\rm horizon}$ is related to the state space of a Chern-Simon theory with 
punctures $j_p$  \cite{Witten,ENP}, with the former arising from the later in the limit of
the CS level $k\to \infty$. Furthermore, the two spaces have precisely the same dimension for any $k$ larger than a given value. 
The space of states of CS theory on a sphere with punctures as describing 
quantum states of black holes has been arrived at in \cite{ENP,lee} using semiclassical considerations based 
on quantizations of theories with boundaries.  

Finally, notice that operator $T_{pp'}$ essentially reads out the ``angle" between the normal to the 
horizon at the punctures  $l_p$ and $l_{p'}$: it can therefore be interpreted as an operator reading 
the extrinsic curvature of the horizon. Thus, the states that are being counted in (\ref{N}) are
those corresponding to different horizon shapes.

\section{Conclusion}

We have given a purely quantum mechanical definition of a black hole as the part of a spin network state
that is not accessible to observables based at infinity.  Semi-classical considerations, such as the analysis of 
boundary conditions at a classical horizon, play no role in this definition.
We have observed that the graph outside the hole, with its intertwiners and spins
is \emph{not} sufficient to describe all degrees of freedom that can
be measured from the \emph{exterior} of the hole.   Additional degrees
of freedom are needed.  We have shown that these additional degrees of freedom
are described by a Hilbert space $\H_{\rm horizon}$, whose elements are
intertwiners between all links puncturing the horizon. This space is related
to the Hilbert space of ${\rm SU}(2)$ Chern-Simon theory with punctures.

Surface states of a black hole are described by Chern-Simon theory also in the analysis 
of \cite{bhe}. The proposal for using ${\rm SU}(2)$ Chern-Simon theory for this 
recently resurfaced in \cite{ENP}, but in fact has a longer history. It was discussed 
in the context of loop quantum gravity by Smolin in \cite{lee}, following earlier suggestions by 
Crane. Here, a related version of this proposal is recovered directly within the full loop quantum gravity. 

We have observed that the operators that read the information in $\H_{\rm horizon}$ are
angle operators between the punctures, and have an intuitive interpretation as measuring 
the extrinsic curvature of the horizon. If the outside geometry is fixed, a black hole is still 
characterized by a number (\ref{N}) of states. These can be seen as describing the ``shape" of the horizon.  
(For a more detailed discussion, see \cite{acta}, and the book \cite{book}.) It is important to emphasize 
that these degrees of freedom are \emph{observable} from the exterior of the black hole: 
if they were not observable they would not contribute to the black hole entropy. Indeed, 
if they had no effect on the external world, and in particular, had no effect on the heat exchanges 
between the hole and the rest of the world, they would not affect the entropy. 

The notion of horizon used here is based on the traditional one
(the boundary of the past of future null-infinity), and it has the same 
limitations.   It would be interesting to find an extension 
of our construction that could capture also the notion of isolated 
horizon \cite{IH}. In this way, in particular, 
one could extend the result presented here also to the scenario where 
information is recovered during, or at the end, of the Hawking evaporation, 
and where, according to the traditional definition, there is no horizon 
\cite{ab}.

We close with a simple comment. If a black hole, seen from its exterior, is described by 
an intertwiner, then an intertwiner can be viewed as a sort of black hole.  This means that a 
semiclassical (``weave") spin network state with Planck-scale intertwiners can be viewed as made 
up with a large number of Planck-scale black holes.   This intuitive image brings
loop quantum gravity closer to Wheeler's initial intuition of a Planck-scale foam.  
At trans-Planckian scale, the quantum energy fluctuations are
such that spacetime disappear into micro-black-holes.  The intertwiners of
the states of loop quantum gravity can be seen as those ``elementary" 
Wheeler's micro-black-holes.

\centerline{--------------}
\vskip.2cm 

This paper is an edited version of notes taken by one of us (CR) in March 98, following a long
discussion with the other author (KK).  At the time, we ended up discarding this idea because the precise 
state counting in \cite{bhe} is based on a ${\rm U}(1)$ CS theory and is \emph{not}  given by (\ref{N}).  
The recent arguments presented in \cite{ENP} renew the relevance in this idea, in
our opinion. CR thank Alejandro Perez for discussing his work before
publication and for numerous inputs on the present paper. KK was supported by an
EPSRC Advanced Fellowship.

\end{document}